\documentclass{iau} 

\usepackage{graphicx}
\usepackage{xcolor}
\usepackage[hyphens]{url}
\usepackage{hyperref}
\hypersetup{
     colorlinks  = true,
     linkcolor   = blue, 
     anchorcolor = BrickRed,
     citecolor   = blue,
     filecolor   = blue,
     menucolor   = blue,
     runcolor    = cyan,
     urlcolor    = blue
}
\usepackage[authoryear]{natbib}

\pubyear{2019}
\volume{355} 
\jname{The Realm of the Low-Surface-Brightness Universe}
\editors{D. Valls-Gabaud, I. Trujillo \& S. Okamoto, eds.}

\title[Utilizing Shell Galaxies] 
{Utilizing Shell Galaxies}

\author[Ebrov\'{a}, B\'{i}lek, \& Jungwiert]   
{Ivana Ebrov\'{a}$^1$, Michal B\'{i}lek$^2$, \and Bruno Jungwiert$^3$}

\affiliation{
$^1$Nicolaus Copernicus Astronomical Center, Polish Academy of Sciences,\\ Bartycka 18, 00-716 Warsaw, Poland; {\tt ebrova.ivana@gmail.com} \\
$^2$Observatoire astronomique de Strasbourg (ObAS), UMR 7550, 67000 Strasbourg, France \\
$^3$Astronomical Institute, Czech Academy of Sciences,\\ Bo\v{c}n\'{i} II 1401/1a, 141 00 Prague, Czech Republic
}

\begin{document}

\maketitle

\begin{abstract}
Stellar shells are low surface brightness features in the form of open, concentric arcs, formed in close-to-radial collisions of galaxies. 
They occur in at least 10\,\% of early-type galaxies and a small portion of spirals and their unique kinematics carry valuable information about the host galaxies. 
We discuss a method using measurements of the number and distribution of shells to estimate the mass distribution of the galaxies and the time since the merger. 
The method is applied on the shells of NGC\,4993 -- a galaxy hosting the electromagnetic counterpart of the gravitational wave event GW170817, to estimate the probable time since the galactic merger.
We used analytical calculations and particle simulations to show that, in special cases, when kinematic data are available, further constraints on mass distribution and merger time can be derived.
Applying the methods to the rapidly growing sample of known shell galaxies will constrain the dark-matter content in the galaxies and reveal detailed information on the recent merger history of the Universe.

\end{abstract}

\firstsection 

\begin{figure}[]
\begin{center}
\includegraphics[trim={0 0 0 9.5cm},clip,width=0.87\linewidth]{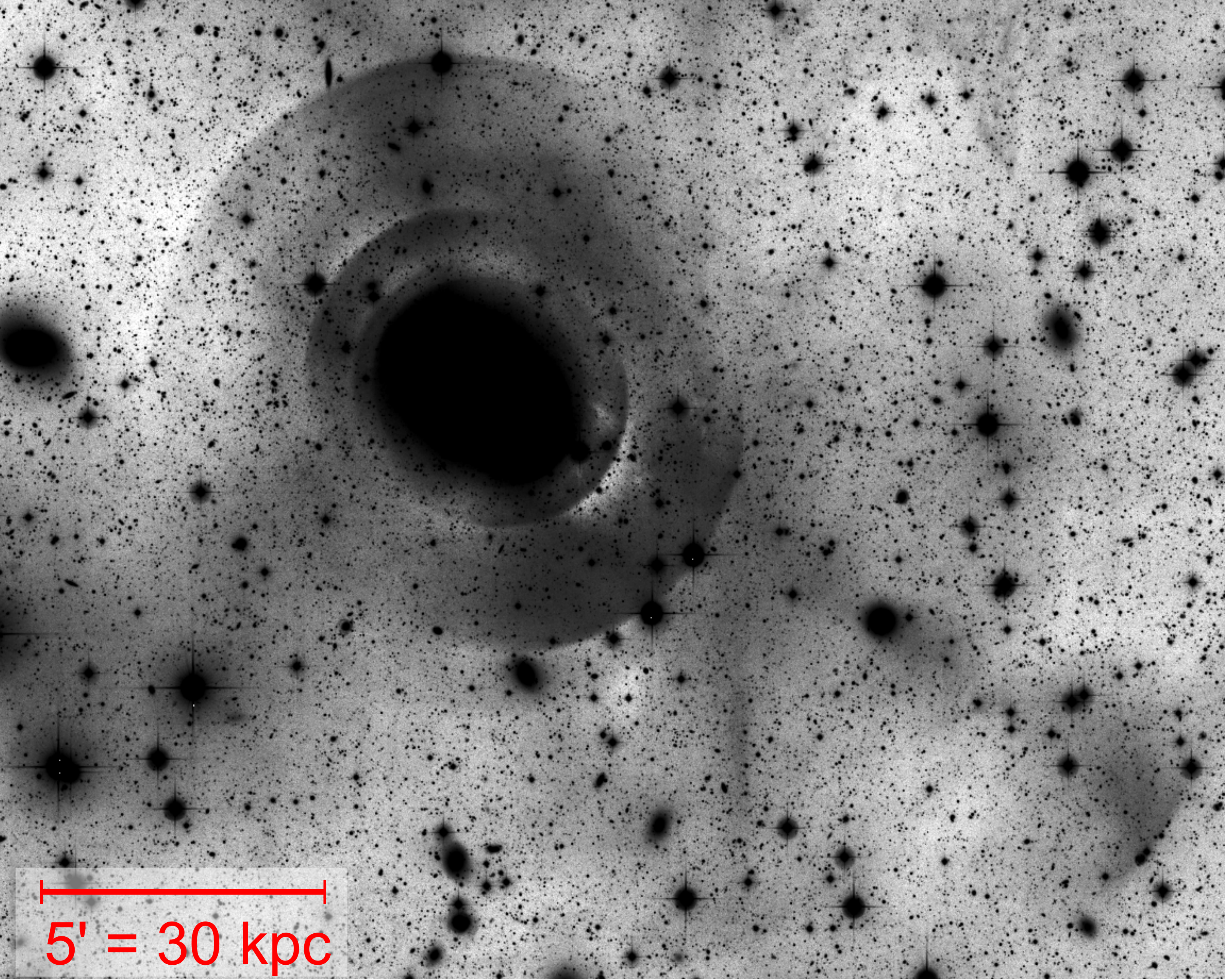}
\caption{
CFHT/MegaCam image of NGC\,3923 with the model of the host galaxy light subtracted, for more details see \cite{bilek16}.
}
\label{fig:3923}
\end{center}
\end{figure}

\section{Shells galaxies}

Shells galaxies account for roughly 10\,\% of all early-type galaxies and they are even more frequent in low density environments and among massive galaxies \citep[e.g.,][]{mc83,tal09,at13}. 
One galaxy can host from one to several tens of shells on galactocentric radii from $\sim$1\,kpc to over 100\,kpc.
There are three types of shell galaxies \citep{pri90,wil87}: Type\,I -- regular shells, interleaved in radius and confined in biconic shape; Type\,II -- arcs randomly distributed all around the galaxy; Type\,III -- irregular.
The shells are made of stars from a galaxy accreted on a close-to-radial trajectory \citep{q84}. 
Traditionally, shells were regarded as minor-merger remnants, but cosmological simulations show that intermediate-mass or even major mergers are likely to produce shells as well \citep{pop18}.
Their unique kinematics, especially for Type\,I shell galaxies (e.g. NGC\,3923, Fig.\,\ref{fig:3923}), enables measurements of the gravitational potential of the host galaxy and the time since the merger.

\begin{figure}[]
\begin{center}
\includegraphics[width=0.85\linewidth]{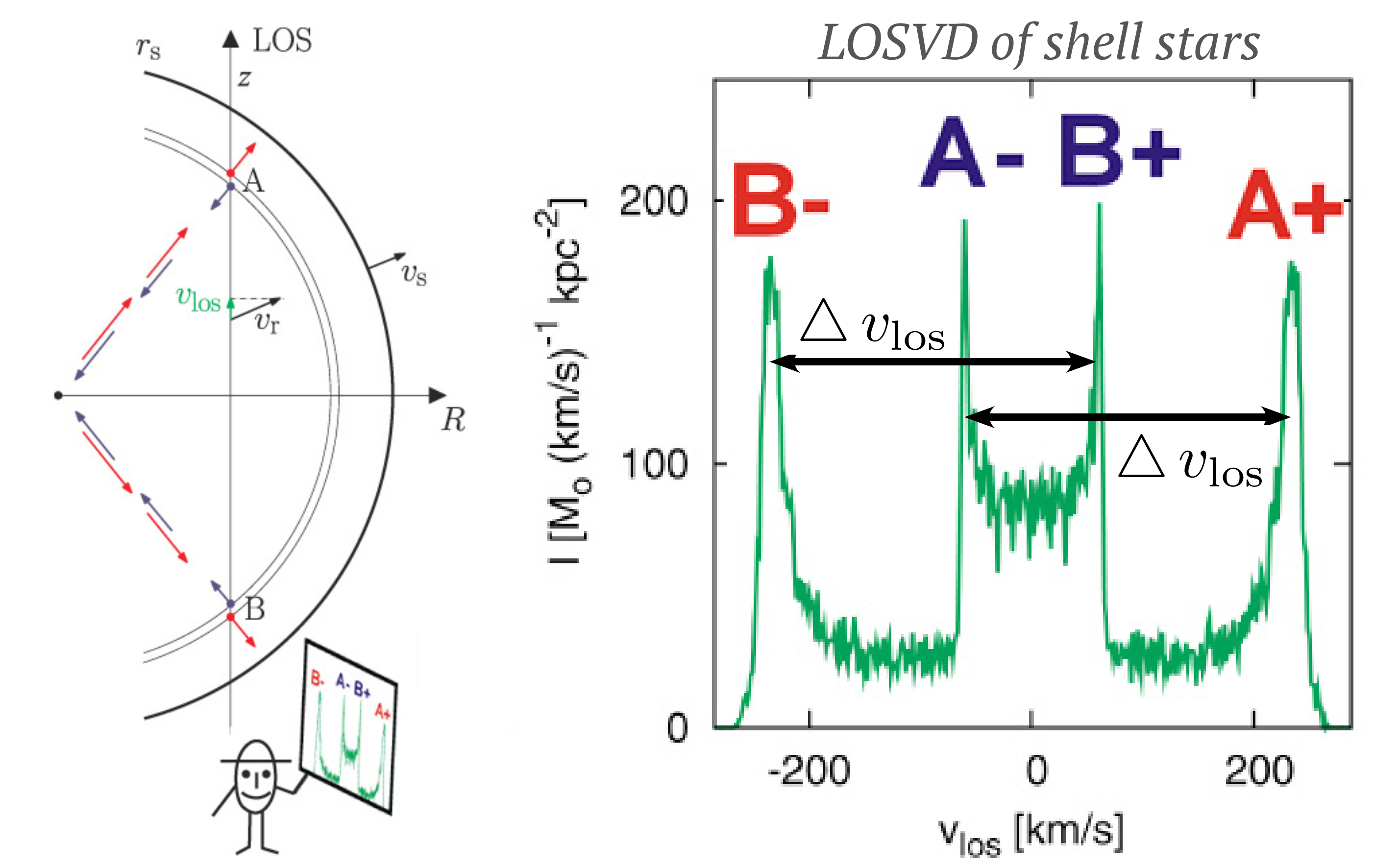}
\caption{
Schema of the genesis of the line-of-sight velocity distribution of shell stars, for more details see \cite{e12sg}.
}
\label{fig:huevo}
\end{center}
\end{figure}

\section{Using shell kinematics}

Measuring the gravitational potential of the galaxy using line-of-sight velocity distribution (LOSVD) of shell stars was first proposed by \cite{mk98} and elaborated in \cite{e12sg}. LOSVD of shell stars shows four distinct maxima, Fig.\,\ref{fig:huevo}, although, in case of shells with narrow opening angles, only two maxima can be present. The value of the circular velocity, $v_{\mathrm{c}}$, in the host galaxy at the shell edge radius is connected to the distance, $\bigtriangleup v_{\mathrm{los}}$, between the peaks of LOSVD, the shell edge radius, $r_{\mathrm{s}}$, and the projected radius of the measurements, $R$, via analytic equation \citep{e12sg}:
\begin{equation}
v_{\mathrm{c}}=\frac{\left|\bigtriangleup v_{\mathrm{los}}\right|}{2\sqrt{\left(1-R/r_{\mathrm{s}}\right)\left[1-4\left(R/r_{\mathrm{s}}\right)^{2}\left(1+R/r_{\mathrm{s}}\right)^{-2}\right]}},
\label{eq:vc,obs}
\end{equation}
or alternatively, using the slope of LOSVD maxima with radius:
\begin{equation}
\frac{\mathrm{d}\!\bigtriangleup\! v_{\mathrm{los}}}{\mathrm{d}R}=-2\frac{v_{\mathrm{c}}}{r_{\mathrm{s}}}.\label{eq:sklon}
\end{equation}
Measuring the kinematics in nearby shell galaxies is still at the frontier of abilities of state-of-the-art telescopes, however attempts have already been done using globular clusters \citep{romanowsky12} or planetary nebulas \citep{lon15} in M87 or red giant branch stars in M31 \citep{fardal12}.

\section{Using shell positions}

Shells can be utilize even in the case when only photometric data is available.
The distribution of shell radii can be used to constrain the gravitational potential of the host galaxy and the time since the merger \citep{q84,dc86,hq87a,hq87b}. 
In \cite{bilek13,bilek14}, we developed a method testing the consistency of a given gravitational potential with the observed shell radii: the time evolution of shell radii in the tested potential is calculated using analytic expressions; if the potential is correct, then the observed shell radii are reproduced by the model at a certain time. In \cite{bilek13}, we verified the consistency of the galaxy NGC\,3923 with MOdified Newtonian Dynamics (MOND), in \cite{bilek14} we used the method to predict a new shell in the galaxy and in \cite{bilek16} we attempted to observe it.

\section{NGC\,4993: the host of GW170817}

\begin{figure}[]
\begin{center}
\includegraphics[width=0.98\linewidth]{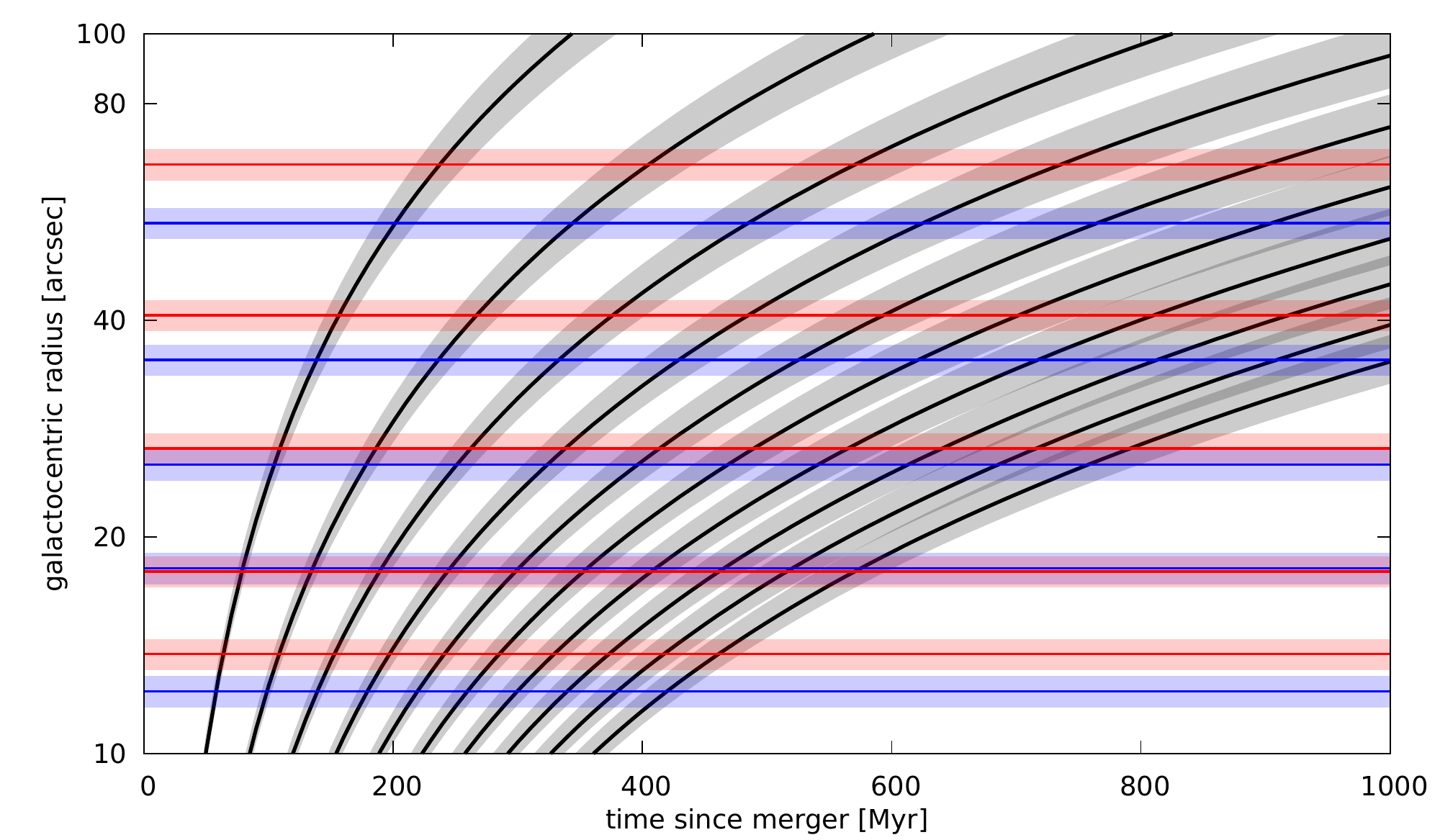}
\caption{
Modeled evolution of shell radii in the assumed gravitational potential of NGC\,4993 (black curves) compared to the
observed shell radii (red/blue lines represent the shells north/south of the galactic centre).
Shaded regions correspond to the uncertainties. For more details see \cite{4993}.
}
\label{fig:radosc}
\end{center}
\end{figure}

NGC\,4993 is the shell galaxy host of the binary neutron star coalescence -- the source of the gravitational-wave event GW170817 \citep{gw17}. The galaxy shows signs that it recently accreted a smaller late-type galaxy and the accreted galaxy is a possible original host of the binary neutron star. In \cite{4993}, we measured the shell distribution using HST/ACS archival data and compares them to theoretical evolution of shell radii in the gravitational potential corresponding to the stellar light and an an adequate dark matter halo.
NGC\,4993, as a Type\,II shell galaxy, does not reproduce the exact positions of the shell radii, Fig.\,\ref{fig:radosc}, but from the outermost observed shell and overall shell number we infer the probable merger time around 400\,Myr. This brings information about the possible age and the birth environment of the binary neutron star.

\section{Conclusions}

Gravitational potential, especially in early-type galaxies, is hard and yet so important to measure, and independent methods to do so are very desirable.
We showed that both photometric and spectroscopic measurements of shell galaxies can be used to estimate the mass distribution of the host galaxy and/or the time since the merger, which is particularly difficult to obtain by other means. 
Applying the methods to the rapidly growing sample of known shell galaxies will constrain the dark-matter content in the galaxies and reveal detailed information on the recent merger history of the Universe.

\acknowledgments
This research was supported by the Polish National Science Centre under grants 2017/26/D/ST9/00449. Based on observations made with the NASA/ESA Hubble Space Telescope, obtained from the data archive at the Space Telescope Science Institute. STScI is operated by the Association of Universities for Research in Astronomy, Inc. under NASA contract NAS 5-26555.

\end{document}